\documentclass[conference]{IEEEtran}
\IEEEoverridecommandlockouts
\usepackage{cite}
\usepackage{amsmath,amssymb,amsfonts}
\usepackage{algorithm}
\usepackage{algorithmic}
\usepackage{graphicx}
\usepackage{textcomp}
\usepackage{xcolor}
\def\BibTeX{{\rm B\kern-.05em{\sc i\kern-.025em b}\kern-.08em
    T\kern-.1667em\lower.7ex\hbox{E}\kern-.125emX}}

\begin{document}


\title{SplitLLM: Collaborative Inference of LLMs for Model Placement and Throughput Optimization}

\author{\IEEEauthorblockN{Akrit Mudvari\textsuperscript{*}\thanks{\textsuperscript{*}Part of work was done while Akrit Mudvari was an intern at NEC Laboratories America.}}
\IEEEauthorblockA{ \textit{Yale University}\\
\textit{NEC Laboratories America}\\
akrit.mudvari@yale.edu}
\and
\IEEEauthorblockN{Yuang Jiang}
\IEEEauthorblockA{\textit{NEC Laboratories America}\\
yuangjiang65@gmail.com}
\and
\IEEEauthorblockN{Leandros Tassiulas}
\IEEEauthorblockA{\textit{Yale University}\\
leandros.tassiulas@yale.edu}
}

\maketitle

\begin{abstract}
Large language models (LLMs) have been a disruptive innovation in recent years, and they play a crucial role in our daily lives due to their ability to understand and generate human-like text. Their capabilities include natural language understanding, information retrieval and search, translation, chatbots, virtual assistance, and many more. However, it is well known that LLMs are massive in terms of the number of parameters. Additionally, the self-attention mechanism in the underlying architecture of LLMs, Transformers, has quadratic complexity in terms of both computation and memory with respect to the input sequence length. For these reasons, LLM inference is resource-intensive, and thus, the throughput of LLM inference is limited, especially for the longer sequences. In this report, we design a collaborative inference architecture between a server and its clients to alleviate the throughput limit. In this design, we consider the available resources on both sides, i.e., the computation  and communication costs. We develop a dynamic programming-based algorithm to optimally allocate computation between the server and the client device to increase the server throughput, while not violating the service level agreement (SLA). We show in the experiments that we are able to efficiently distribute the workload allowing for roughly $1/3$ reduction in the server workload, while achieving $19$ percent improvement over a greedy method. As a result, we are able to demonstrate that, in an environment with different types of LLM inference requests, the throughput of the server is improved.  
\end{abstract}


\section{Introduction}

LLMs, in recent years, have been playing a progressively more transformative role in our lives by enhancing Natural Language Processing (NLP), exemplified by the Generative Pre-trained Transformers (GPTs) \cite{radford2018improving}. These language processing tools have served as a backbone for various applications including chatbots, virtual assistants, translation services, and improved search engines. In recent years, it could be said that this technology has played one of the most profound roles in changing how human beings interact with technology. 

This fame of LLMs has been primarily driven by general language models, such as GPT-4, which has a tremendous capacity to generate new text based on the aggregated knowledge acquired through training on vast and multidisciplinary data, and Bidirectional Encoder Representations from Transformers (BERT) \cite{devlin2018bert}, which has been used for a wide array of downstream tasks, including text classification. However, there are now various LLMs, which serve a wide array of tasks based on language processing, transcending numerous domains, languages, or specific tasks. For instance, legalBERT \cite{chalkidis2020legal} is a domain-specific LLM providing legal document analysis, contract review, and more, while BioBERT \cite{lee2020biobert} focuses on biomedical and clinical texts, providing question-answering services within this domain. Other LLMs focus on multilingual and translation services, which include M2M-100 used for providing high-quality translation services without using English as an intermediary, or DeepL Translator \cite{deepl2024}. Furthermore, LLMs may be specifically designed for specific tasks. For instance,  RoBERTa for NER (Named Entity Recognition), which is a model fine-tuned over RoBERTa, helps identify and classify entities such as names of people, organizations, or locations, from a text. Besides, Wav2Vec \cite{schneider2019wav2vec} is a task-specific tool specializing in converting audio signal to discernible text. Furthermore, there have been numerous visual transformers, such as CMT \cite{guo2021cmt}, for image classification. 

At the core of these LLMs are the Transformers \cite{vaswani2017attention}, which are a type of neural network architecture that utilizes multi-headed self-attention mechanism as the core method of recognizing the importance of the relationship between different representations (i.e., different words) in large and complex learning environments. In LLMs, the most popular and widely used application of Transformers, this mechanism captures the relationship between different words in a sentence. Other aspects of this architecture include positional encoders that assign a position-aware identity to the input tokens (parts of each input), and other general Neural Network (NN) structures such as feed-forward NNs. Most of the deep learning models in the past have relied heavily on layers with linear computational and memory complexity with respect to input size; examples include feed forwards NNs, Convolutional Neural Networks (CNNs) or Recurrent Neural Networks (RNNs). However, Transformer models have computational and memory complexity that increase quadratically as the input sequence lengths increase. As we will discuss in further detail in section \ref{subsection:examinations}, the quadratic complexity means that for longer language inputs, the computation and memory costs grow very quickly. Beyond this, the fact that these LLMs have parameter size ranging from hundreds of millions to tens of billions means that the computation and memory costs are very high even for the LLMs with smaller input sequence lengths. 

The Transformer-driven models, while highly costly and only computationally feasible in powerful data center settings in many cases, are nonetheless highly desirable services at the edge of the network. A lot of other non-LLM, AI-driven services are already being deployed at the edge of the network, in Internet of Things (IoT) devices, AR/VR devices, cellphones, and more. This has been facilitated by increased bandwidth, improved latency, and better reliability. Furthermore, network slicing and mobile edge computing are poised to facilitate a wider range of services towards the edge of the network. So, the demand for such models to be deployed for inference is set to grow, and as a result, we can expect an ever-expanding demand for these LLM services at the network edge in the coming years \cite{de2023chatgpt}. This will, in turn, put tremendous pressure on the service providers who need to commercially afford the limited capacity under the edge-cloud paradigm. This may be done through another vendor or self-owned servers. Besides the costs, this is set to put pressure on the servers, especially with fluctuating demands leading to throughput issues. The promising capability of these LLMs to provide user-friendly AI for various tasks has been realized, and combined with the idea of multi-modality \cite{huyen2023multimodal}, different types or modules of LLMs may need to be deployed at different locations within the network.   

Current efforts for reducing computational loads, especially for the longer sequences in Transformers, focus on reducing the computational complexity of the self-attention layers \cite{beltagy2020longformer,chen2021scatterbrain,zaheer2020big}.  Primarily, this involves approximation of the sparse, low rank, or some combination thereof, of those two approaches. The primary idea is to find computationally and memory-wise cheaper methods of calculating the approximations of the $n \times n$ attention matrix, usually opting for computations that are less than quadratic in complexity. In the sparse approximation approaches \cite{tay2020sinkhorn}, the most relevant elements within the matrix are computed while ignoring the parts that are not likely to be as significant. In the low rank approximation approaches, lower-rank representation of the attention matrices are calculated. Some other methods \cite{zaheer2020big} attempt to calculate a combination of those different approaches. While useful, most of these methods fail to reach the accuracy of a full attention matrix, and the ones that do end up approaching the capabilities of the full attention matrices will have lost the computational improvements \cite{tay2020long}. While these approaches allow for the utilization of fewer GPU resources and for a shorter time (for either inference or training), it is not always adequate for a vast majority of cases to run most of the LLMs outside the powerful servers, especially if accuracy needs to be optimized. 

In recent years, there has been a tremendous growth in facilities that afford computational services under the fog computing paradigm. On the cloud front, there have been numerous efforts leading to improvements in the latency and bandwidth of the services provided by the data centers. For instance, the first half of 2023 saw the largest amount of data center construction in the history \cite{cbre_data_center_trends_2023}, and the data center colocation market is set to reach 131.80 billion dollars, a significant rise from 2022's 57 billion dollars \cite{techjury_data_center_stats}. As the data centers become more prevalent, there is a much better chance of better bandwidth provisioning and lowered latency due to geographical proximity. With the proliferation of 5G cellular networks across the world and the inclusion of Mobile Edge Computing (MEC) as a core technology under 3GPPP standardization \cite{3gpp}, we can expect to find more servers closer to the user/data. Numerous efforts have been made towards scalable and efficient provisioning of different services under such cellular networks \cite{mudvariMLscale21, 9477278, mudvariMLscale22}, and many services can benefit from being deployed towards the network edge \cite{8486021}. The concept of edge computers serving the users closer to the end devices has moved forward significantly in the commercial domains, and the edge data center market, which is already valued at 11.01 billion dollars, is set to grow by 6 times within the next decade \cite{precedence_edge_data_center}. As a result, we will soon find a pervasive network of computational resources, allowing for low latency, and high bandwidth services for AI and other applications. For instance, a user may be able to acquire services from MEC servers close to a base station, as well as closely located powerful data centers with numerous state-of-the-art GPU resources. 

The current approach of running LLMs and other large models involves offloading the raw data to a server location where the entire inference is completed. This approach has apparent problems that make it a less than ideal approach, including higher network costs \cite{mudvari2023adaptivecompress, eshratifar2019bottlenet}, potential throughput loss due to server congestion, and privacy infringement due to offloading of raw input data \cite{jeong2018computation, yang2022differentially, yao2022privacy}. On the other hand, processing data in the local devices, even the ones that are moderately capable, would pose significant issues including a very high task completion latency due to a lack of computational capability. As discussed earlier, this would in effect be significantly worse as the sequence length increases. Besides the discussion earlier, in section \ref{subsection:effective} , we will discuss in detail how longer sequence length and self-attention layers make it very difficult to run LLM inference in weaker devices. Hence, a solution in the form of a distributed intelligence approach is split inference \cite{mudvari2023adaptivecompress, kang2017neurosurgeon}. In this approach, certain layers are computed locally on a client device, and then the remaining layers are forwarded to the server devices for further inference. Split Inference (at least layer-wise splitting as we define and discuss in this paper) refers to splitting a neural network into multiple partitions so that different layers of the deep learning model can be computed in different ways. For instance, consider a neural network with layers $l_1,….,l_m,….,l_n$ where $m<n$; here split learning can involve processing $l_1,….,l_m$ locally and then offloading $l_{m+1},….,l_n$ to a server for further computations. Such split learning methods can provide us with various benefits over simply offloading the model as a whole. First, unlike offloading raw data, we can send a processed output, which is able to remain privacy preserving \cite{jeong2018computation, wadhwa2023pfsl, yang2022differentially, thapa2021advancements, yao2022privacy}. Second, the granularity provided by such a division of tasks allows for more optimal scheduling and placement. Third, as evidenced by recent research works \cite{wadhwa2023pfsl, arivazhagan2019federated}, it could also help with personalization of learned models by keeping the locally trained parts more personalized and having a globally shared set of layers for aggregated learning. And finally, the more important benefit we will explore in this work is an intelligent offloading under the split inference paradigm, where the goal will be to reduce the computational load on the server by optimally delegating certain computational load to the end devices, leading to a throughput improvement for the servers handling large number of AI demand. Layer-wise split learning differs from feature-wise split learning, where in feature-wise split learning, different portions of the input data are distributed across multiple compute nodes for parallel training.    

In this work, we develop an intelligent splitting algorithm that will leverage the properties of the LLM models and the input sequence properties (i.e., length) to develop an efficient resource allocation method, which will be optimal under practical assumptions pervasive in today's fog/edge-cloud computing paradigms. Our goal will be to reduce any computational load (i.e., computational cost in FLOP or GPU memory) in a task-constrained server so that the throughput is improved. The formulation will demand strict user requirements in the form of task completion latency, and we will demonstrate and discuss the effectiveness of this approach for different situations, including for different LLM models and bandwidth availabilities.     

\section{Related Work}

In this section, we discuss different works that are of relevance to our topic and contrast those with the novelty of our work. 

Numerous works in literature have focused on scheduling and placement of computational workloads in a distributed setting; however, a multitude of work have also been conducted on extending these problems to an edge-cloud architecture. For instance, in \cite{poularakis2020service}, the number of requests is maximized, with storage, computation, and communication costs as constraints, and in \cite{poularakis2020approximation} similar problem is solved for data-intensive requests. Similarly in  \cite{pasteris2019service}, a service placement algorithm with a near-optimal solution guarantee is presented. Other works, including \cite{sonkoly2021survey, malazi2022dynamic, luo2021resource}, similarly work on different types of placement and scheduling methods for implementing state-of-the-art techniques for efficient allocation of resources under the edge-cloud paradigm. While these works focus on general problems, taking into account the unique characteristics of deep learning architecture can help improve the efficiency of scheduling and placement decisions in multiple ways. For instance, different factors such as energy consumption and network utilization are used in \cite{murthy2016deep} to decide which deep learning models should be run and ``where'' within a network. In another work, \cite{han2016mcdnn}, scheduling is developed with approximate models, where some accuracy is traded away for guaranteed service, under constraints such as device energy storage, cloud computing costs and capacity, and execution deadlines. An online variant of the solution is also developed in this work to tackle real-time workloads.

Complementary to scheduling or placement solutions, some methods help reduce the workload while processing deep learning architectures, by employing methods that could be envisioned for different deep learning architectures. Such methods include quantization of the entire network \cite{han2015deep}; in the quantization approach, parameters or variables, during inference or backpropagation, can be stored in a way that saves memory and/or computation costs. A much more specific and strict version of quantization is binarization \cite{rastegari2016xnor}, where the model is stored and processed in a binary form. Other methods that aim to reduce the model size include pruning, which includes pruning the parameter space of the model \cite{molchanov2019importance,carreira2018learning, zhao2019variational} or pruning the feature space of the model \cite{hu2016network, he2017channel, peng2019collaborative}. These different methods, while effective, inevitably compromise the performance guarantee, since the process often involves sacrificing some model accuracy for an increased compute performance. Even in rare cases where performance may not fall significantly (sometimes, slight pruning may even help tackle over-fitting and increase accuracy for test data), the guarantee of performance cannot be given. Our major contribution in this paper is on developing a method, that not only ensures continued guarantee of optimal performance but also works in a way where these different methods can be complementary to our implementation.  

There have been various works in the literature where distributed or split implementations of neural network models aim for collaborative intelligence under different objectives. In \cite{kang2017neurosurgeon}, the authors proposed a method for collaborative intelligence between end devices and the mobile edge. They divided the model for partial computation at each end, aiming to find the optimal point for splitting the model to improve inference latency and energy efficiency. In \cite{samikwa2022ares}, the authors develop a method focusing on splitting deep learning architectures at an optimal location to share the model between a client and a server, intending to accelerate the training time, minimizing the effects of bottleneck client devices, and reducing the energy consumption. And in \cite{samikwa2023disnet}, layer-wise split learning is combined with feature-wise splitting to accelerate inference time. A multitude of similar works exist, but the goal of optimization tends to be energy or resource reduction at the client devices with no concern for server throughput, and splitting only takes place at one defined location.    

The splitting learning paradigm has since been implemented in various works in literature, including for splitting transformer models for different goals. For instance, in \cite{he2022destr}, the transformer model for classification and box regression are split and computed separately towards an object detection goal. In \cite{park2021federated}, the split learning is implemented for a visual transformer with feature extraction and classification done at the client, and the rest of the task offloaded to the server. In \cite{patel2023splitwise}, a split learning paradigm is implemented, but the neural network architecture itself isn't split. Here, the prompt phase is run on the client device, and the token generation phase is run on the server. Our work can be considered complementary to this kind of approach, in that we provide further granularity for splitting within a single forward pass. These past works do not implement decision-making for layer-wise split deep learning models in multiple places, which would allow us to harness the power of high-bandwidth next-generation networks and the varying computational properties of different model layers.

Several studies have aimed to enhance the efficiency of Transformers in particular, especially for processing longer inputs, to reduce the extensively large computation and memory consumption. Longformer~\cite{beltagy2020longformer} combines windowed self-attention and global attention to sparsify the full attention matrix. Similarly, Bigbird~\cite{zaheer2020big} introduces a sparse attention method that incorporates random, windowed, and global attention, demonstrating improved performance in tasks such as question answering and summarization. Sparse sinkhorn attention~\cite{tay2020sinkhorn} and Reformer~\cite{kitaev2020reformer} incorporate learnable patterns into the attention module. Vyas, Katharopoulos, and Fleuret~\cite{vyas2020fast} propose clustered attention, computing attention only for centroids in clustered queries. Other works focus on kernel-based and feature mapping methods, like Performer~\cite{choromanski2020rethinking}, Reformer~\cite{kitaev2020reformer}, and Linformer~\cite{wang2020linformer}, which improve self-attention efficiency through grouping, clustering, or designing fixed sparse patterns, albeit at the cost of expressiveness. In contrast, our work focuses on a totally different angle: instead of decreasing the computation costs and time for the models at the expense of performance, we consider the split of computation to increase the throughput of LLM inference at the server.

{\bf Novelty:} In this work, we develop a method of collaborative inference through model splitting for transformer-driven architectures, where the splitting decision is efficient and takes into consideration both the computation and the communication resources. Such an approach makes it suitable for the edge-cloud networks of today and the near future, where significant benefits from resource management could be realized. We develop a method that guarantees optimal results, beyond most of the methods described above, and on top of that, our method is complementary to most of those methods; our method can incorporate various methods that aim to minimize computation or communication costs, and we discuss briefly how our efficient splitting scheme can work alongside aforementioned sparse and low-rank self-attention approximation approaches. We test our method for different LLMs and a visual transformer, comparing it against a greedy implementation. This helps establish the efficacy of our method across different sequence lengths, model types, network environments, and more. Finally, we demonstrate that the method will be useful in improving the throughput at the server that is tasked with providing computation resources to different clients demanding different types of LLM inference services.

This paper is arranged in the following way. In section \ref{section:method}, we formulate the split decision-making problem, describe our novel algorithm for intelligent split decision-making, and prove its optimality. In section \ref{section:evals}, we begin by demonstrating that efficient splitting can help reduce the computation load at the server without violating the latency requirement. We also show that such efficient splitting schemes are compatible with other methods of computation load reduction such as sparse and low rank approximation of the self attention matrix. Then we show that our optimal splitting algorithm can outperform a greedy method at reducing the server load, followed by the demonstration that this leads to an improvement in throughput at the servers. WE conclude the work in section \ref{section:conclusion}.

%





\section{Method}
\label{section:method}

In this section, we begin with problem formulation that allows us to establish the relationship between a server and its clients within a communication network, and model the deep learning inference with split/collaborative execution. Within this section, subsection \ref{subsection:formulation} describes the problem formulation that captures the networked infrastructure, alongside the optimization goals. In subsection \ref{subsection:algorithm}, the modeled problem is analyzed, computational complexities are discussed, and an optimal algorithm for solving the formulation is presented in subsection \ref{subsection:formulation}. Finally, in subsection \ref{subsection:proof}, the optimality of the algorithm developed is proven in section \ref{subsection:algorithm}. 

\subsection{Problem Formulation}
\label{subsection:formulation}

Let us consider a network with the server $s$ providing placement service for the end user $e \in E$. Let the model being used for inference be $m \in M$. Our goal is to find the optimal splitting decision $\pi(s,e,m)$, which ensures that a minimal amount of the task is offloaded to the server, while also satisfying a certain latency constraint $\Lambda(e)$, as per the user application requirement. For the connection between the devices $e$ and $s$, we will define a bandwidth $\delta(s,e)$ to denote the download rate (data going from the device $s$ to device $e$) and a bandwidth $\delta(e,s)$ to denote the upload rate (data going from the device $e$ to $s$). This deep learning model can be considered to be made up of layers $l \in L(m)$, with each layer $l$ taking a computation time of $c(e)_l$ to process in device $e$ and  $c(s)_l$ to process in device $s$. For the sake of simplicity during the discussion of the method, we will refer to the added latency of computation for the layer $l$ while moving the task from $s$ to $e$ as $c_l$. Let the input tensor for each layer be of size $\tau_l$, then the download time can be obtained as $d_l = \tau_l / \delta(s,e)$ and upload time is given by $u_l = \tau_l / \delta(e,s)$. 

In a distributed placement scenario, a split neural network model may have certain layers allocated to the server, while the remaining layers are computed locally. The optimal placement policy that aims to minimize the processing load on the server would try to offload certain tasks towards the end device without violating the service level agreement, i.e., the latency requirement of the given application. The decision of whether or not to run a particular layer on the end device is represented by a binary decision variable $x_l$, where a value of $1$ represents that the task can be run in the end device, and a value of $0$ represents that the tasks must be run on the server. We will also introduce $l(prev)$ to represent the previous layer of layer $l$. 

Then, the latency constraint for the given model can be defined as: 
\begin{equation}
    \label{eqn:latencyConstraint}
    \begin{split}
    \Lambda(m) \geq \sum_{l \in L(m)} { x_l(c(e)_l + (1-x_{l(prev)}) d_l) }
    \\ + 
    \sum_{l \in L(m)} { (1-x_l) (c(s)_l + x_{l(prev)} u_l) } 
    \end{split}
\end{equation}

In equation \ref{eqn:latencyConstraint}, the first half of the sum is for situations where the computation for any layer $l$ takes place at the end device, i.e., when $x_l$ is $1$, with the computation time being $c(e)_l$, and the download time  $d_l$ is considered only when the previous layer was in the server (when $x_{l(prev)}=1$). The second half is similar but accounts for the situations where the computation takes place at the server. 

As long as the task is completed within $\Lambda(m)$, the quality of service agreement is expected to be satisfied, so the goal is then to minimize the number of tasks offloaded to the server. For each layer $l \in L(m)$, we will define the computation load of that layer to be $r_l$. Then the optimization goal is given by
\begin{equation}
    \label{eqn:loadMin}
    \min_{x_l \forall l \in L(m)} {(1-x_l) r_l }
\end{equation}
$r_l$ is a parameter that may be collected in multiple ways such as by sampling certain metrics during a sample inference. For example, GPU memory usage can be calculated by monitoring the GPU (for instance, we used Nvidia's system management interface to monitor GPU memory usage). We primarily decided to select the FLOP (Floating Point Operations) for each layer to measure the computation costs. We obtained the FLOP values by calculating them for each layer, and then verified our measurements using an open-source tool, fvcore \cite{fvcore2024}. The developed algorithm however is designed to work for any layer-wise calculated metric that we aim to minimize. 

\subsection{Problem Analysis and Algorithm}
\label{subsection:algorithm}

\begin{figure}[]
  \centerline{
    \includegraphics[width=0.4\textwidth]{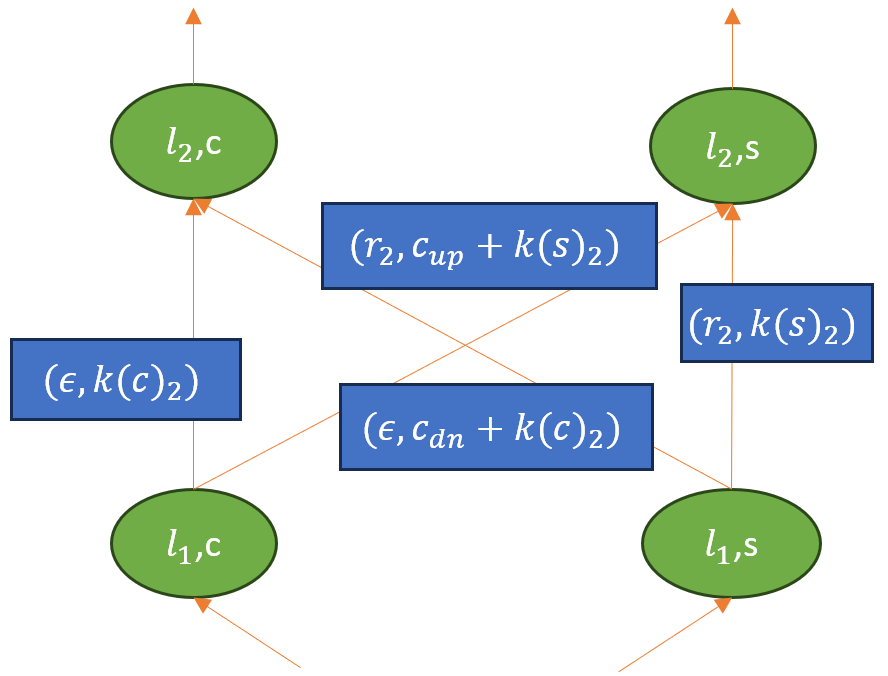}
  }
  \caption{Illustration of problem as a constrained shortest path problem} 
  \label{fig:constrained_dijkstra}
\end{figure}

Towards solving the aforementioned formulation, the first step is to realize that the problem may not be trivial. Consider the neural network above to be a graph $G$ with each $(layer,location)$ pair as a node in the graph. A snippet of the equivalent graphical representation of this model inference is illustrated in Figure \ref{fig:constrained_dijkstra}. Running layer $l_1$ in the end device $e$ is a node $(l_1,e)$, and $(l_1,s)$ is another node representing running the same layer in server $s$ instead. A link represents the communication between two subsequent layers. For instance, if after layer $l_1$ is computed in server $s$, it may be further computed in $s$ as well such that the next hop is $(l_2,s)$, or it may be downloaded to the client such that the next hop is $(l_2,c)$. In a case where the computation occurs in the server again or at the server twice in a row, the link between nodes is a tuple of the form $(r_2,k(s)_2)$. Here, the first part of the tuple represents the ``weight" of the link, which is the resource expended (which we are trying to minimize), and since the next layer is run in the server, this cost is $r_2$. The second element of these link/tuple represents the ``cost" of the link, which is how much time was expended in communication, as represented by $k(s)_2$. If the decision was to download instead, the ``cost" would entail both compute time at the client ($k(c)_2$) as well as a download cost ($k_{dn}$), giving a total cost of $k(c)_2+k_{dn}$. But since we don't care much about resource usage at the client, this can be represented by a small cost value $\epsilon$. Thus, downloading from the server to the client would give the link a tuple representation $(\epsilon,k(c)_2+k_{dn})$. 

Any inference task must begin with $l_1$ (starting location may be $s$ or $e$), and end at $l_L$ (at one of the devices). For instance, suppose we are constrained by the fact that inference starts at $(l_1,e)$ and ends at $(l_L,s)$. Then our task is to reduce the weight at the server as denoted by equation \ref{eqn:loadMin}, which means minimizing the computation cost at the server/minimizing traveling through nodes that represent computations in the server, i.e., nodes $(l,s)\ \forall l \in L$. 
But, traversing the graph also has a cost constraint: each link in the graph represents a certain time delay, and the sum of these delays as we traverse the links (from start to end node) must not exceed the latency constraint as mentioned in equation \ref{eqn:latencyConstraint}. 

Here, we have reduced the problem formulation from earlier into a constrained shortest path routing problem, which is a class of NP-hard problems \cite{pugliese2013survey}. As it stands, this makes it very hard for our optimization goal to be resolved efficiently and on time. However, we are able to develop a pseudopolynomial algorithm to solve this problem by relying on certain relaxations that are unlikely to significantly change the optimal solution. Our solution utilizes an approach where the problem can be solved with a complexity of $\mathcal{O}(Lw)$, where $L$ is the number of distinct layers in the deep learning model, and $w$ is an ``integerized" representation of latency limit $\Lambda(m)$. By this we mean that we select $w$ to be an integer value: for instance, for a deadline of 500 milliseconds, it could be 500 with the smallest unit 1 representing 1 millisecond. Without this method, the computational complexity of solving the model would be $\mathcal{O}(2^n)$. As most transformer-driven neural networks, but also most other neural networks, are composed of numerous layers, the complexity makes it very hard to apply a brute force approach to obtain the results quickly. 

\begin{algorithm}[h!]
    \caption{Algorithm for obtaining layer placement policy}
    \label{algorithm: networked_placemment}
    
    \textbf{Input: } For each layer $l \in L$, client inference time $i_l \in \hat{I}$, input download time $d_l\in \hat{D}$, input upload time $u_l \in \hat{U}$, and computation resource usage/cost $r_l$. Inference deadline $\Lambda$, Smallest time unit $T$, start-at-client flag $SaC$\\
    \textbf{Output: } Layer schedule policy $\pi$ \\
    
    \begin{algorithmic}[1]
    \STATE Initialize: Layer schedule Vector $\pi$ of size $|L|$  with $\pi_l = 0 $ $ \forall l \in \{0,...,|L|\}$
    \STATE Obtain: ``Integer Equivalent" for every time-related inputs with the function $I,D,U,W = Inteq(\hat{I},\hat{D},\hat{U},\Lambda,T)$ (algorithm \ref{algorithm: joint_algorithm})
    \STATE Initialize: Storage matrices $C$ and $S$ of size $(|L|+1,W+1)$ with $C(k,j)=0$ and $S(k,j)=0$ $ \forall k \in \{0,....|L|+1\} $ and $ \forall j \in \{ 0,....,W+1 \} $ 
    \STATE Initialize: $\epsilon=-W$
    
    \STATE Trivial case: first row of $C$ and $S$ 
    
    \FOR{rows $k = 2,....,|L|+1$}
        \FOR{columns $j = 1,....,W$}
            \IF {$j-u_k < 0$}
                \STATE Set: $c2s = \epsilon$
            \ELSE
                \STATE Set: $c2s = j-u_k$
            \ENDIF
            \STATE Set: $c2c = max(0,j-i_k)$
            \STATE Set: $s2c = max(0,j-i_k-d_k)$
            \STATE Set: $C(k,j) = max(C(k-1,c2c),S(k-1,s2c))$ 
            \STATE Set: $S(k,j) = max(C(k-1,c2s),S(k-1,j))$ 
        \ENDFOR
    \ENDFOR
    \STATE Initialize: $w=W$ to represent time resource remaining 
    \FOR{entry $k =|L|,....,1$}
        \STATE Set: $c2c = max(0,w-i_{k})$
        \STATE Set: $c2s = max(0,w-u_{k})$
        \STATE Set: $s2c = max(0,w-i_{k}-d_{k})$
        \STATE Set: $s2c = 0$
        \IF {$\pi[k+1] == 0$}
            \IF {$S(k,s2s) < C(k,s2c)$}
                \STATE Set: $\pi[k] = 1$
                \STATE Set: $w = w - s2c$
            \ELSE
                \STATE Set: $w = w - s2s$
            \ENDIF
        \ELSE
            \IF {$S(k,c2s) < C(k,c2c)$}
                \STATE Set: $\pi[k] = 1$
                \STATE Set: $w = w - c2c$
            \ELSE
                \STATE Set: $w = w - s2c$
            \ENDIF
        \ENDIF

    \ENDFOR
    \end{algorithmic}
\end{algorithm}

\begin{algorithm}
    \caption{Algorithm for obtaining integer approximation (Inteq function)}
    \label{algorithm: joint_algorithm}
    
    \textbf{Input: } For each layer $l \in L$, client inference time $i_l \in \hat{I}$, input download time $d_l\in \hat{D}$, input upload time $u_l \in \hat{U}$, inference deadline $\Lambda$, and smallest time unit $T$ \\
    \textbf{Output: } Integer approximation $W,I,D,R$ for inputs $\hat{I},\hat{D},\hat{U}, \Lambda$\\
    
    \begin{algorithmic}[1]
     
    \FOR{rows $k = 1,....,|L|$}
        \STATE Set: $i_k = round(\hat{i}_k / w)$
        \STATE Set: $d_k = round(\hat{d}_k / w)$
        \STATE Set: $u_k = round(\hat{u}_k / w)$
    \ENDFOR
    \STATE Set: $W = round(\Lambda / w)$
    \end{algorithmic}
\end{algorithm}

Our algorithm with a computational complexity of $\mathcal{O}(lw)$ is shown as Algorithm \ref{algorithm: networked_placemment}. Towards the decision-making, for each layer $l \in L$, model information to be provided are client's inference time $i_l \in \hat{I}$, download time for the output of the previous layer from server to client $d_l \in \hat{D}$, upload time for the output of the previous layer from client to server $u_l \in \hat{U}$, and resource cost as explained earlier $r_l$. other non-layer specific information to be provided are the inference deadline for the model $\Lambda$, and the smallest time unit $T$. Since our algorithm requires the aforementioned time units to be treated as integers, $T$ is a variable that decides the smallest possible unit $w$. It was seen that a very small value of $T$ was still sufficient to run the algorithm in real time, and this value can effectively be 1~ms or even lower. The algorithm will return as output the layer scheduling policy $\pi$ of size $|L|$. This is a binary vector with $\pi_l=0$ if the algorithm decides that the layer should be run in the server, and $\pi_l=1$ if that part should run locally.

In line 2 of algorithm \ref{algorithm: networked_placemment}, the information ${\hat{I}, \hat{D}, \hat{R}, \Lambda, T}$ are taken as input and using algorithm \ref{algorithm: joint_algorithm} (also called Inteq), the integer equivalent $I,D,U,W$ as discussed earlier is obtained. The dynamic programming aspect of the algorithm starts at line 3 with the creation of matrices $C$ and $S$ of size $(|L|+1,W+1)$ and all values initialized to zero. The two tables are dynamically updated with $C(l,w)$ referring to optimal placement up to capacity $w \in W$ for up to layer $l$ at the client with $l_{th}$ layer remaining at the client, and  $S(l,w)$ referring to optimal placement up to capacity $w \in W$ for up to layer $l$ at the server with the $l_th$ layer remaining at the server. The lowest possible value, $\epsilon$ needs to be sufficiently small, and this value can be less than $-W$. 

Lines 6-18 in algorithm \ref{algorithm: networked_placemment} refer to the population of the tables $C$ and $S$, which will be used by the selection part of the algorithm (lines 19 onward) for returning the optimal placement decision variable $\pi$. An explanation of how this section works would benefit from looking at an arbitrary layer $l$ and weight $w$. At that point $(l,w)$, 4 different actions are possible: server-to-client $s2c$ (where the previous layer was computed on the server and the current will be computed on the client), server-to-server $s2s$ (both previous and current layers are computed on the server), client-to-client $c2c$ (both previous and current layers are computed on the client), and client-to-server $c2s$ (where the previous layer was computed on the client and the current layer will be computed on the server). The time cost associated with each of those configurations is calculated accordingly: for instance, $c2c$ would only have the cost of computing in the $k_{th}$ layer, since staying in the same device does not incur any communication cost. 

It is necessary to ensure that the costs for the $j_{th}$ column of the matrices $C$ and $S$ only consider feasible actions, so the cases that will cause these costs to exceed $j$ will only have a value of $\epsilon$. For instance, $s2c$ will need to have a total cost value of $i_k+d_k$ (for downloading to the client and then processing in the client), and this value should be less than or equal to the availability value $j$. Also, line 8 ensures that enough budget is always available, to ensure that there are enough resources to upload to the server; this is done by assigning a very small value $\epsilon$ to $c2s$ when uploading is infeasible. The algorithm automatically ignores policies that have such large negative values.  Finally, once all the associated costs are calculated for up to a given $j$ and $k$, then
values of $C(k,j)$ and $S(k,j)$ are now available, and these values help determine the cost associated with having $k$ layers under cost/time budget $j$, at client or server device respectively. The tables, starting from the top-left are filled for each row first, left to right, before moving to the row below, as suggested by liens 6-7, until the entire table is filled. 

\begin{figure}[]
  \centerline{
    \includegraphics[width=0.4\textwidth]{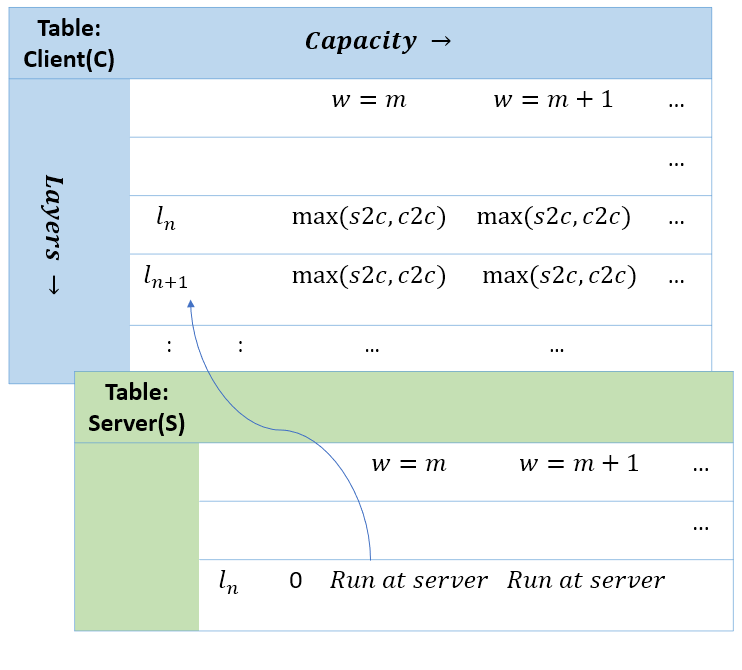}
  }
  \caption{Illustration of dynamic allocation tables as described in Algorithm \ref{algorithm: networked_placemment}} 
  \label{fig:dynamic_table}
\end{figure}

Figure \ref{fig:dynamic_table} intuits the process of assigning values to the table. For instance, let us assume the task were optimal/assigned at $l_{n-1}$ in the server $S$ (Just to illustrate what happens when optimal assignment is in $S$ at $l_n$). Then the options are for the tasks to be offloaded to the client $C$ for processing at layer $l_{n+1}$, with the cost $s2c$ described earlier, or continue running in the server. For the client $C$, the optimal decision at layer $l_{n+1}$ is the better option between fetching data and processing the output of $l_n$ at $S$, or continuing with the policy of running at client, hence $max(s2c,c2c)$.   

The second part of algorithm \ref{algorithm: networked_placemment}, lines 19-39, are for using the now populated tables $C$ and $S$ to discern the best policy $\pi$, used for assigning each layer to either the server or the client. We begin by assigning the value $W$ to a variable $w$, which will keep track of how much time resource has been spent. The goal is to keep track of integerized variable $w$ ($0 \leq w \leq W$) that has been assigned to the layers for which a decision has already been made. We begin iterating bottom-to-top across matrices $C$ and $S$, starting with the decision for the last layer and ending with the first. For any layer $k$, the decision to whether or not to keep the task at the client is decided by the entry in $\pi[k]$. The first step (lines 21-24) within this part of the algorithm is for deciding what was the cost of reaching the selected state (either reaching client or server) for $k$ layer is, given the different costs associated with the processing layer and transferring the layer output if upload/download is needed. Take line 21 assuming $w > i_{k}$, where $w$ represents the time resource remaining and $i_{k}$ is the cost of processing the data in the client device in $i^{th}$ layer. The value of $c2c$ then represents how much budget remains after the decision to run $k^{th}$ layer at the client device uses up, given that $k^{th}$ layer was run on the client as well. Similarly, $c2s$ represents the same for when $k^{th}$ layer was run on the client and $k+1^{th}$ on the server. $s2c$ and $c2s$ are calculated similarly.       

The next step is to determine the allocation of $k^{th}$ layer given the values of $k+1^{th}$ layer, and the associated costs of processing and transferring. If $k+1^{th}$ layer was to be processed in the client as decided by the decision variable $\pi[k+1]$ being $0$ (line 25), then the next step would be to compare the costs to determine whether the given step would be optimally allocated at the server ($\pi[k]=1$) or the client ($\pi[k]=0$). This decision is made by comparing the values at $S(k,s2s)$ and $C(k,s2c)$ to determine which of the two allocations would have left the most resources for the future layers. After a decision is made, it is important to remove the resources used up by the $k^{th}$ layer. For instance, in line 28, by which point it will have been decided that $k+1^{th}$ layer was going to be processed in the server, and $kth$ layer was going to be run in the client. Then, the new weight is given by $w - s2c$, where $s2c$ is the cost running $k^{th}$ layer on the client given by $i_k$ added to the cost of sending data from client to server $d_{k}$.  

Upon the completion of algorithm \ref{algorithm: networked_placemment}, the optimal placement policy $\pi$ is obtained.  

\subsection{Proof of Optimality}
\label{subsection:proof}

While the problem formulation was done with our specific application requirement, with each layer having only two possible states, ``on the server" or ``on the client", it is actually possible to extend the method to situations where each layer's computation can be done in several possible different ways or states. This generalized approach is proven below, with ``2 possible states" being a specific case for our algorithm. 

Let's consider a Directed Acyclic Graph (DAG) $G=(V,E)$. Let each vertex $v \in V$ have certain weight $w_v$, and let each edge $e(v_1,v_2) \in E$ from vertex $v_1$ to vertex $v_2$ have a weight $w_e(v_1,v_2)$. In this directed graph $G$, consider a source node $v_s$ and a destination node $v_d$. The value/reward obtained by visiting any node $v_i$ is given by $r_i$. Consider the dynamic programming approach, and consider a node $v_i$ with all the nodes leading to $v_i$ given by $v \in V_b(i)$. 

In the dynamic programming approach, for each node, $v_i$, we assign a vector $S_i$ of size $W$ such that $S_i(w) \ \forall w \in W$ gives the best value for the assignment constraint of size $w$, which means that for a budget of $w$, the value stored will be the best possible one. 

This implies that in the DAG, each of the possible neighbors $v_k \in V_b(i)$ is also assigned a vector $S_k$. So the value for $V_i(w)$ is given by:
\begin{equation}
    V_i(w) = \max {(r_i + V_k(w-w_e(v_k,v_i)\  \forall v_k \in V_b(i))}
    \label{eqn:bestVal}
\end{equation}

To show that this method always generates the optimal result, we proceed with proof by induction. So we begin with the first entry where $i=1$. In this case, for any $w \in W$, the best result is always achieved with the possible inclusion of $v_1$ if possible, since no other nodes are considered.

Next, it should be shown that if the optimal solutions are provided for $ \forall v_k \in V_b(i)$, then the optimal solution also exists for $v_i$ following equation \ref{eqn:bestVal}. Let this best value be obtained by reaching node $v_i$ through node $v_{\hat{k}}$; then the maximum value is $r_i + V_{\hat{k}}(w-w_e(v_{\hat{k}},v_i)$, where $r_i$ is a constant for any $w$, and by assumption for $ \forall V_k(w)$, $V_{\hat{k}}$ must be an optimal value.

And since $v_d$ is an arbitrary node in the DAG, this must also be true for the destination node.


\section{Discussion and Evaluation}
\label{section:evals}

In this section, we evaluate the results for the method discussed in section \ref{section:method}. We start with subsection \ref{subsection:examinations}, where we use a sample case to demonstrate that intelligent splitting decisions can help reduce the server workload more efficiently, i.e., without causing a significant delay in end-to-end latency. In subsection \ref{subsection:sparselow}, we demonstrate how efficient splitting could also be beneficial when the attention layers are approximated with low-rank or sparse representations (however, we do not consider these methods during the examinations in the subsequent sections, since our primary goal is to design a method that provides a performance guarantee). Then in subsection \ref{subsection:effective} we discuss, by analyzing different types of transformer models, how the efficient splitting method we developed can help reduce server load while respecting the task completion deadline. We will demonstrate the efficiency of the method across different network and model features, including different sequence lengths of the input data, latency requirements, and bandwidth availability. Finally in subsection \ref{subsection:throughputs}, we show that the developed method is capable of improving throughput for the servers providing the resources for different types LLM inference, under different SLAs and network conditions.

\subsection{Examinations for Efficient Splitting}
\label{subsection:examinations}

\begin{figure}[]
  \centerline{
    \includegraphics[width=0.4\textwidth]{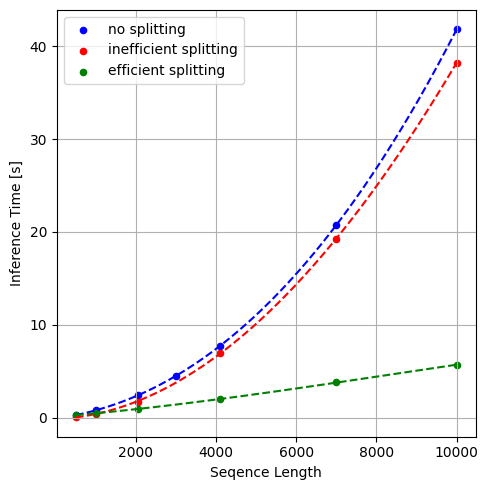}
  }
  \caption{Inference time under different split policies} 
  \label{fig:times_across_policies}
\end{figure}

\begin{figure}[]
  \centerline{
    \includegraphics[width=0.4\textwidth]{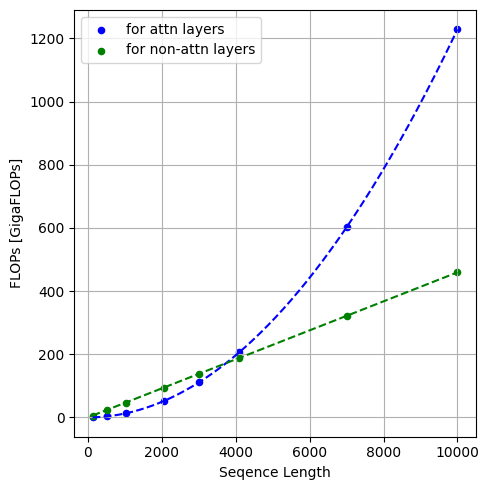}
  }
  \caption{FLOP used by different types of layers} 
  \label{fig:flop_across_policies}
\end{figure}

\begin{figure}[]
  \centerline{
    \includegraphics[width=0.4\textwidth]{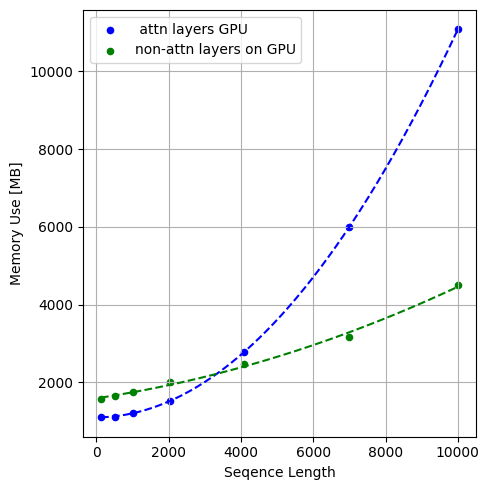}
  }
  \caption{Memory usage by different types of layers} 
  \label{fig:mem_across_policies}
\end{figure}

In a split learning or inference model, if each of the layers were similar in terms of computation requirements, or in terms of each layer's output size, then there would not be a need to invest resources towards optimal splitting policy, and there would be no need to split multiple times either. However, as we have discussed earlier, the uneven nature of the steps involved in the inference of the LLMs, as well as the need for network efficiency during communication, means that an efficient splitting can help reduce the computation and communication workloads and lead to throughput improvements on the server side. To demonstrate the scope of efficient decision-making during spit inference, we begin with a simple setup and some heuristic demonstrations. 

We begin by discussing an experiment that shows the promise of efficient splitting for transformer-driven models by deploying a model with 12 attention layers among other components such as positional encoders and classifier (as used in BERT \cite{devlin2018bert}, or ALBERT \cite{lan2019albert} architecture). For this task, we set up a distributed learning environment with RTX 3090 as the server with GPU, and a resource-constrained 1 CPU core (limited using taskset) as the client. In this split paradigm, the model is deployed with PyTorch, and TCP socket programming allows for communication between the server and the client. Then, we observe the inference across different input sequence lengths, $s$, with different splitting schemes, including a ``no splitting" scenario where all computations are carried on the weak end device, a somewhat ``efficient splitting" scenario where only the attention tasks are handled by the server/GPU, and ``inefficient splitting" where all tasks except attention layers are handled by the server. Each results are obtained by running the inference 5 times and taking an average. It must be noted that the models aren't trained for all the demonstrated sequence lengths, since in most cases larger sequence lengths lead to very high learning time, inference time, and resource costs (something we partially attempt to alleviate with this work). One of the objectives of this work is to develop a framework for larger input sequence lengths as well, which would allow for inference over larger inputs in future research and commercial endeavors. 

The results of this experiment are shown in figure \ref{fig:times_across_policies}. ``no splitting" in the figure represents a scenario where no splitting happens and everything is run on the client, while ``inefficient" represents a very inefficient scenario where only the computationally intensive attention layers are run on the client. Finally, ``efficient splitting" represents a scenario where only the attention layers are run on the server. As we observe in the figure, the quadratic nature of self-attention layers has a significantly adverse effect on the inference completion time, especially as the sequence length increases. As a result, it becomes much more preferable to run certain layers on a powerful server where a significant reduction of computation load can be achieved through powerful GPUs. 

In Figure \ref{fig:flop_across_policies}, we see the relative FLOP (Floating Point Operations) allocated to different types (attention vs. non-attention) of layers across different sequence lengths. These values were analyzed for different types of attention, feed-forward, and other layers mathematically, but also verified later using fvcore \cite{fvcore2024} library. It can be seen that as the input data size grows, the number of FLOP allocated to these different types of layers behave differently; the other layers show a linear growth, while the attention layers have the FLOP values increasing quadratically. Let us begin by considering a situation where $s=4000$ in figures \ref{fig:times_across_policies} and \ref{fig:flop_across_policies}. Here, while roughly similar computational cost is assigned to either the attention or the non-attention layers (figure \ref{fig:flop_across_policies}), the inference completion times are very different for two the different splitting schemes. For instance, the latency requirements can be 4 times more relaxed / higher in efficient splitting vs. inefficient splitting while a similar resource-saving is achieved. This effect can be observed across different optimization objectives. In figure \ref{fig:mem_across_policies}, we observe the same resource-saving potential for GPU memory allocation for inference tasks, where attention layers occupy slightly more GPU memory (roughly 1.2x) at $s=4000$, but the inference time is 4 times less as before. Hence, it becomes obvious that relatively more or less efficient policies exist for different optimization goals, further validating the need for formulation and algorithm in section \ref{section:method}. In this section, we will rely on FLOP count for analysis, but the formulation and algorithm are completely agnostic to which computation resource is selected for minimization.      


\begin{figure}[]
  \centerline{
    \includegraphics[width=0.4\textwidth]{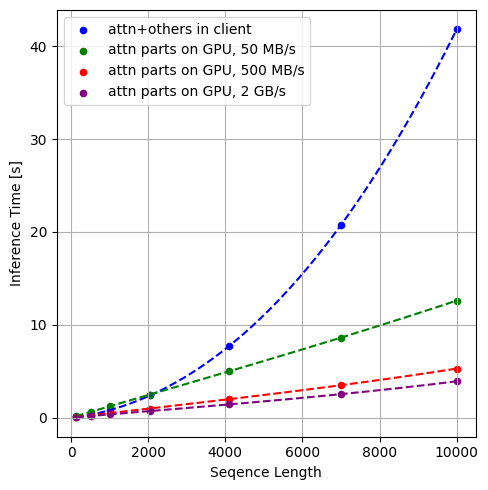}
  }
  \caption{Inference time under different split policies and bandwidths} 
  \label{fig:times_across_policies_bandwidths}
\end{figure}

To ascertain the efficiency of the method, one of the earlier tests conducted was to see how the method could work for different bandwidth availability. In figure \ref{fig:times_across_policies_bandwidths}, we can observe that the benefit of efficient splitting discussed above could be observed across different bandwidths, and the improvement is more pronounced as we increase the available bandwidth. The improvement is seen across the different input sequence lengths, as the green, red, and purple lines are significantly below the blue curve, especially as the sequence length increases. Another point to ascertain would be the assumption that server implementation was significantly faster than client implementation. Our implementation showed that running the entire model with sequence length 4096 would take 7.727 seconds in the client, while it would only take 0.0979 seconds in the server, verifying the assumption that completion time is much faster on the server side.

\subsection{Sparse and Low Rank Approximation}
\label{subsection:sparselow}

\begin{figure}[]
  \centerline{
    \includegraphics[width=0.4\textwidth]{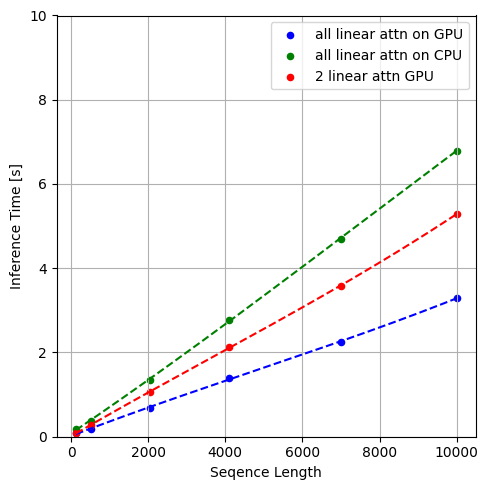}
  }
  \caption{Inference time for different splitting policies when linear approximation of attention matrix is used.} 
  \label{fig:diff_linear_policies}
\end{figure}

\begin{figure}[]
  \centerline{
    \includegraphics[width=0.4\textwidth]{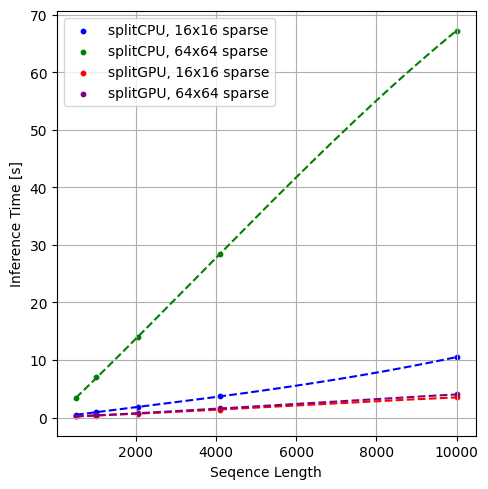}
  }
  \caption{Inference time for different splitting policies when sparse approximation of attention matrix is used.} 
  \label{fig:diff_sparse_policies}
\end{figure}

While the goal of this paper is to demonstrate the benefits of collaborative inference without a loss in optimal performance, the formulation could be extended to situations where approximate models that trade off accuracy for better inference time and lower computation cost are deployed. This includes models with sparse or low-rank representations of the self-attention matrix. As a proof-of-concept, we demonstrate how efficient policies could be formulated in such scenarios. In figure \ref{fig:diff_linear_policies}, we refer to a solution, i.e., \cite{chen2021scatterbrain}, where a linear combination of low-rank matrices helps approximate the full attention layer. As we see, for different sequence lengths, depending on the latency requirements, all, none, or $2/3$ of the linear layers could be run on the GPU. Similarly, in figure \ref{fig:diff_sparse_policies}, we can see that different sparse approximations (with different smaller matrix sizes \cite{zaheer2020big}) can be used to achieve different collaborative inference results. Here, splitCPU refers to transmitting attention layers to another machine without GPU, while splitGPU refers to transmitting attention layers to another device with GPU; the size, i.e., $16\times16$, is the size of the smaller matrices used to approximate full attention matrix.      

\subsection{Effectiveness of the Method}
\label{subsection:effective}

In this section, our primary objective is to demonstrate the efficiency of our method, based around algorithm \ref{algorithm: networked_placemment}, for efficiently optimizing the load reduction objective as formulated in equation \ref{eqn:loadMin}, while ensuring that the end-to-end inference latency constraint is always satisfied. We begin with an experimental setup as described in the preceding section \ref{subsection:effective} with a couple of changes. First, we are not limited to one model but experiment across different models, including multiple LLMs and an image-recognition transformer. We simulate the inter-device communication this time, where instead of using a socket programming approach, the communication delay and transmission delay are simulated, which greatly saves time in running inference across different bandwidths, models, and more. We opt for this approach since the efficacy of the approach is already established in the earlier subsection. 

We compare our algorithm against the 'Greedy' approach, where the computation is greedily assigned to the client device for the first layers, so long as the latency constraint allows it. This is based on the split decision-making as described in \cite{kang2017neurosurgeon}. Solutions that stem from such splitting, including \cite{shao2020bottlenet++, eshratifar2019bottlenet, patel2023splitwise}, are complementary to our method. Greedy splitting is proposed as the offline solution in \cite{shi2021dnn} for a non-server-client non-LLM-specific paradigm.     

Before running inference on the models, the proof of optimality presented in section \ref{subsection:proof} was complemented with numerical verification. Random number generators were used to demonstrate that the dynamic programming approach is optimal, and the same random number generator was used to show that the algorithm performs better than the 'Greedy' approach. This was done by running the numerical calculations multiple times. The greedy approach entails computing as many layers as possible on the client side, but the selection mechanism is greedy, in that the layers that come earlier are always selected until the budget runs out. Then the remaining layers have to be processed on the server side. 

Before trying different models, we begin with a model as designed in \cite{vaswani2017attention}, which is a transformer with 6 encoder and 6 decoder layers leading to 18 self-attention layers, plus the adjacent feed-forward NNs and other layers, alongside the positional encoding layers, classifier and more. As before, we run the inference 5 times and take the average to get inference data. We also select different bandwidths and assign latency so that the latencies roughly fall in the range where roughly all, to almost none, of the data can be offloaded to the client devices. Each subsequent latency is half of the larger one, and this way the data are not unfairly selected. Similarly, different transmission rates are selected as well, and a communication delay of 10~ms is added. It must be noted that such splitting paradigms are suitable under different environments including the fog paradigm with fairly decent communication resources. In the communication infrastructures of the past such as 4G cellular networks, such learning paradigms would not be as suitable since significant network resources are consumed.

\begin{figure*}[ht!]
  \centering
    \includegraphics[width=\textwidth]{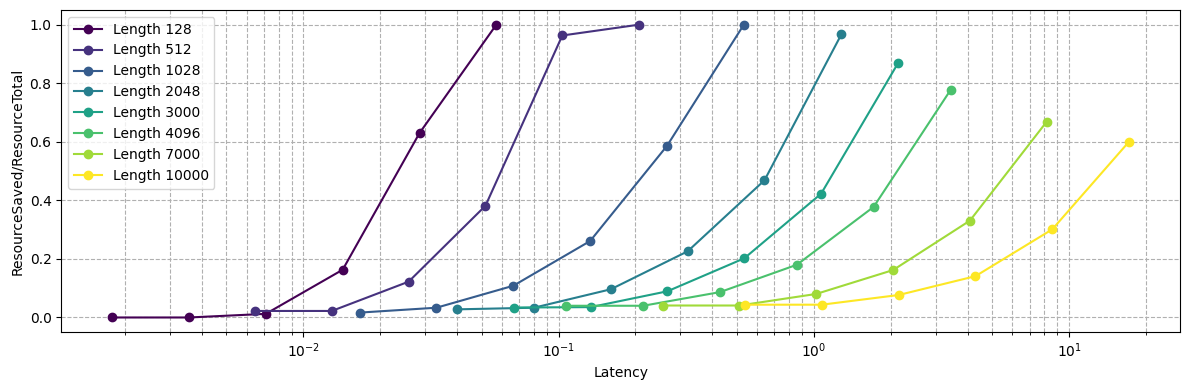}
  \caption{Resource usage across different latency requirements}
  \label{fig:Savedvslats}
\end{figure*}

\begin{figure*}[ht!]
  \centering
    \includegraphics[width=\textwidth]
    {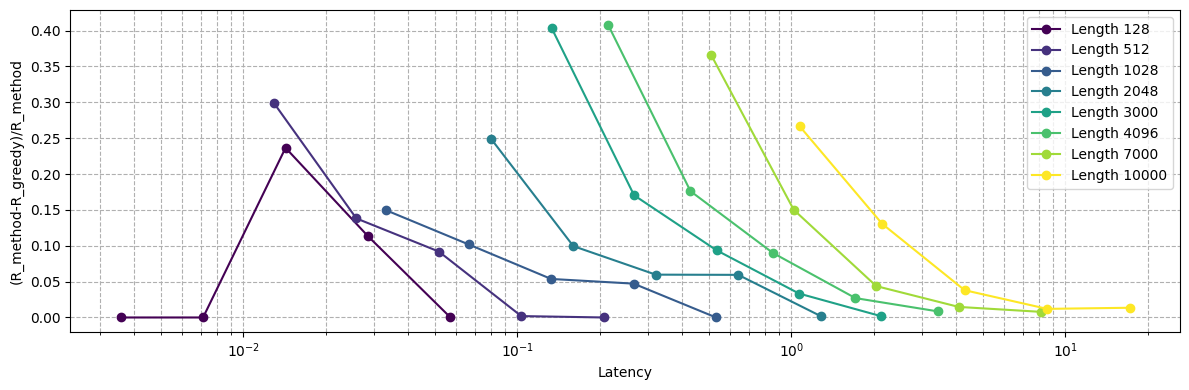
    }
  \caption{Improvement over greedy method across different latency requirements}
  \label{fig:vsGreedyvslats}
\end{figure*}

In figure \ref{fig:Savedvslats}, we observe the amount of computation offloaded to the server; While these values differ across different latencies, sequence lengths, and different model settings, the average percent of resources offloaded to the server was 28.9 percent. With this ratio of inference tasks offloaded to the server, the improvement of our method over the greedy method was found to be 14.6 percent. In figure \ref{fig:vsGreedyvslats}, we can see the improved performance over the greedy approach for most of the values across the sequence lengths and the latencies. These values, across latencies and sequence lengths, were averaged over different bandwidth availability. As is intuitive, we can observe in figures \ref{fig:Savedvslats} and \ref{fig:vsGreedyvslats} that as the latency becomes high enough for most of the tasks to be offloaded to the client, the benefit over the greedy approach diminishes; the benefit is more pronounced for stricter latencies. 

\begin{figure}[ht!]
  \centering
    \includegraphics[width=0.4\textwidth]{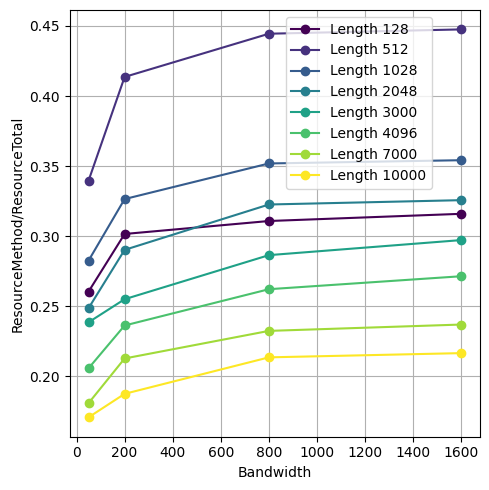}
  \caption{Resource usage across different bandwidth availability}
  \label{fig:BWsavedvslats}
\end{figure}

\begin{figure}[ht!]
  \centering
    \includegraphics[width=0.4\textwidth]{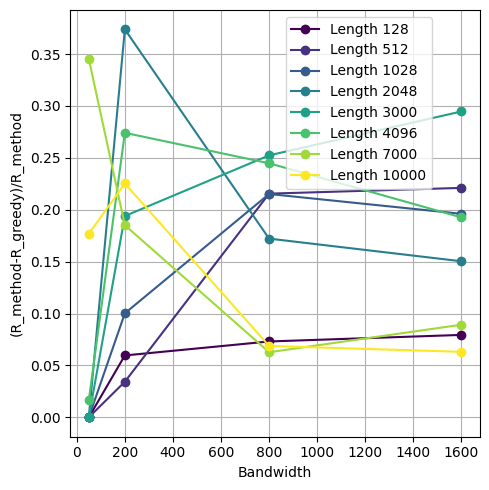}
  \caption{Improvement over greedy method across different bandwidth availability}
  \label{fig:BWvsGreedyvslats}
\end{figure}

In figure \ref{fig:BWsavedvslats}, we see the average tasks offloaded to the server across different bandwidths, and as is intuitive, at higher bandwidths, it becomes easier to offload more tasks to the server, which is seen with the growth in the amount of resources that can be offloaded with the increasing bandwidth availability. In figure \ref{fig:BWvsGreedyvslats}, we see the improvement of our method over the greedy approach, in reducing workload at the server, when compared across different transmission rates.

Next, we repeated the experiment for two other layers, i.e., BERT (12 encoder layers, so 12 attention layers) and GPT-2-like model (24 attention layers). In each of the cases, we allowed for roughly a bit under 1/3rd of the resources to be offloaded to the clients on average (27.8 percent for BERT, and 29.2 percent for GPT-2-like model). We mention ``GPT-2-like" since the exact parameter space for GPT-2 isn't fully public knowledge. For the case of BERT, the improvement over the greedy approach at efficiently offloading to the client was found to be 5.5 percent, while it was found to be 12.5 percent for the GPT-2-like model. This showed us that the improvement across different environments and model configurations, which we saw earlier, was also observed across different models. While it was not always the case, the fact that we generally saw improved performance for larger sequence lengths, as well as larger models, suggests that models with larger computational requirement/parameter size might see a bigger improvement. 

Until this point, and the core objective of the paper, focuses on the language models, and the behavior of the language models across different environments or model designs. While this analysis is not considered in other sections, we also attempted to understand the efficiency of our method on a transformer that specialized in visual recognition \cite{guo2021cmt}. While we start with the default Imagenet \cite{imagenet_cvpr09} images of size 224x224, we scale it, just for inference tests, up to 4 times, to study the effect of input size on the splitting policy. We recognize that unlike the language models, where a longer input size is highly desirable and different efforts have been made towards achieving this, a larger input size isn't as sought after in the domain of visual deep learning methods. So we do not use these results during the throughput consideration (section \ref{subsection:throughputs}). Unlike language models, visual transformers are significantly deeper and the NN has a much more varying structure. Here, we found that with 44.3 percent network resource saved, the improvement over the greedy approach was 55.4 percent. Since the input sizes fluctuate a lot layer-after-layer in such visual transformers, and since the greedy algorithm needs to reserve certain time resources for uploading in the worst-case situation where the time deadline may come to an end while processing is still in the client device and output of the layer is large, the performance for the greedy method was worse for the visual transformer as opposed to the language models. Since our method guarantees optimality under the given relaxations, such situations are not of concern to our algorithm.   

\subsection{Throughput Improvement}
\label{subsection:throughputs}

In the earlier section, \ref{subsection:effective}, we demonstrated that our method can lead to an efficient split inference such that the server load is minimized given the latency requirement. In this section, we will demonstrate how these reductions can lead to efficient throughput improvements during the dynamic scheduling of model inference requests. In this section, we will use the data collected across different models, bandwidth requirements, and latencies to simulate a random arrival process with an inter-arrival rate $\beta$. The arriving traffic is served by a server with a computational capacity described by $\Omega$. The tasks that do not have enough resources provided upon arrival will stay in a queue that has a ``first-in-first-out" principle, and the resource availability is frequently checked to see if the next request can be executed at the time. The running time of the tasks is based on the task completion deadline and the frequency of executions (some tasks may be asked to be executed multiple times, i.e., up to 10 times). Since this demonstration relies on high arrival rates and provisioning of the services to a large number of requests, we rely on simulations to demonstrate the improvements achieved from our method. 

\begin{figure}[ht!]
  \centering
    \includegraphics[width=0.4\textwidth]{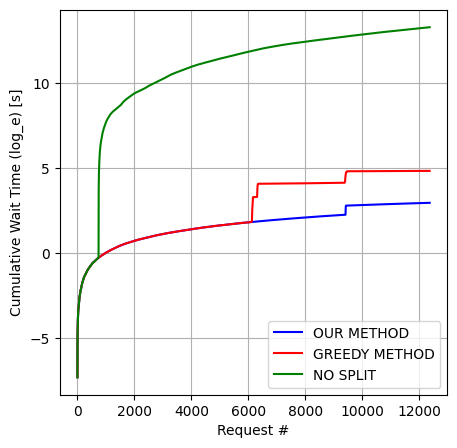}
  \caption{cumulative wait time across different methods, $\beta=57/1000$}
  \label{fig:throughputOne}
\end{figure}

\begin{figure}[ht!]
  \centering
    \includegraphics[width=0.4\textwidth]{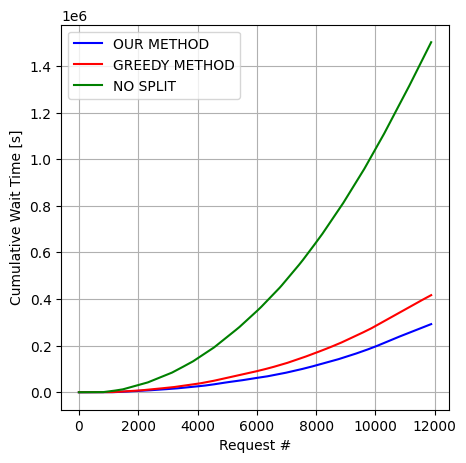}
  \caption{cumulative wait time across different methods, $\beta=45/1000$}
  \label{fig:throughputTwo}
\end{figure}

In figure \ref{fig:throughputOne}, we observe the throughput when $\beta=57/1000$. Capacity $\Omega$ here is described as being able to serve 500 requests on average (pre-calculated) at a given time, and the total number of requests served is 14,949. Here, we observed that the maximum wait time for our method was 1.36~s, while it was 3.13~s for the greedy method, and 110.62~s for the no-split method. The average wait times were 0.0061~s for our method, 0.0682~s for the greedy method, and 47.507~s for the no-split method. This showed a significant improvement when our method is employed. For a different value $\beta=45/1000$, we observe that our method has a higher cumulative queue wait time than before, but it is significantly lower than no-split or greedy approaches, as shown in figure \ref{fig:throughputTwo}. Here, as the arrival rate for the requests was increased and all methods faced a queue, the maximum delay grew to 59.2~s for our method, 87.2~s for the greedy method, and 270.5~s for the no split approach. This is a kind of delay that would violate service agreements in many cases, but it can be observed that our method still has a significantly better performance. On the other direction, if the rate was increased to $\beta=60/1000$ for instance, the average wait time would be negligible for our method or greedy approach, while it would be 73.67~s for no split approach.    

The goal of this section was to show how efficient splitting policies can help with the improvement of throughput at the servers providing such LLM-based services. Needless to say, the server capacity should be designed to handle expected traffic in a data-centric way, but with our method, it can be observed that there is a better throughput performance across different workloads for a given server capacity. Such designs should be data-driven and well-planned, but a method that reduces the server load at the individual request level, such as ours, is bound to improve the performance in cumulative deployment scenarios.  

\section{Conclusion}
\label{section:conclusion}

As the LLMs have proven to be an extremely pervasive technology, new methods are needed to tackle the high costs of computation and resource congestion that such a growth is likely to pose. Recognizing that this issue will become progressively worse in the near future, we developed a collaborative inference scheme that exploits the nature of transformer models and provides an efficient splitting algorithm for the LLMs with different input sequence lengths, model types, network settings, and other requirements. We show that our method outperforms a greedy approach by $19$ percent on average, towards decreasing computation costs at the server by roughly $1/3^{rd}$. We also show that this improvement in turn increases throughput at the server. 

\bibliographystyle{IEEEtran}
\bibliography{main}

\end{document}